# Modeling of jitter correction methods for asynchronous optical sampling

**Mayuri Nakagawa,[1,2,*] Natsuki Kanda,[1,3] and Ryusuke Matsunaga,[1]**
1. The Institute for Solid State Physics, The University of Tokyo, 5-1-5 Kashiwanoha, Kashiwa, Chiba 277-8581, Japan
2. Network Service Systems Labs., NTT Inc., 3-9-11 Midori-cho, Musashino, Tokyo, 180-8585, Japan
3. RIKEN Center for Advanced Photonics, RIKEN, 2-1 Hirosawa, Wako, Saitama, 351-0198, Japan

*Corresponding author: mayuri.nakagawa@ntt.com

**Abstract:** Timing jitter management is crucial in asynchronous optical sampling (ASOPS) and dual-comb spectroscopy to maintain spectral bandwidth and fidelity. However, a quantitative framework predicting jitter correction efficiency from laser phase noise has been lacking. Here, we present formulation and simulation models that evaluate jitter-suppression dynamics based on phase noise spectra. Our model quantifies the contrast between conventional triggering and software-based jitter correction. It also enables retrieval of a jitter-free spectrum by compensating for jitter-induced power degradation. This work provides a universal design framework for optimizing diverse ASOPS systems by predicting residual jitter prior to system construction.

## 1. Introduction

Asynchronous optical sampling (ASOPS) and dual-comb spectroscopy (DCS) have attracted considerable attention as efficient spectroscopic techniques for achieving high frequency resolution and high scan rate in terahertz time-domain spectroscopy [1–4], pump-probe measurements [5,6], and molecular spectroscopy [7–9]. These methods employ dual femtosecond lasers with slightly different repetition frequencies $f_{\mathrm{r1}}$ and $f_{\mathrm{r2}}$. While ASOPS directly reconstructs time-domain waveforms, DCS retrieves optical spectra through multiheterodyne detection [10–12]. Assuming that laser #2 is sampled by laser #1, the sampling rate is defined by $f_{\mathrm{r1}}$, and the pulse delay increment $\left(\frac{1}{f_{\mathrm{r1}}}\right)-\left(\frac{1}{f_{\mathrm{r2}}}\right)=\frac{\Delta f_{\mathrm{r}}}{f_{\mathrm{r1}}f_{\mathrm{r2}}}$ gives the effective time resolution, where $\Delta f_{\mathrm{r}}\equiv f_{\mathrm{r2}}-f_{\mathrm{r1}}$. Therefore, $f_{\mathrm{r1}}\cdot\left(\frac{\Delta f_{\mathrm{r}}}{f_{\mathrm{r1}}f_{\mathrm{r2}}}\right)=\frac{\Delta f_{\mathrm{r}}}{f_{\mathrm{r2}}}$ gives the down-conversion ratio under that assumption. Because the available time window is limited to one pulse interval, the achievable frequency resolution is $f_{\mathrm{r2}}$. Consequently, the scan rate for this window becomes $f_{\mathrm{r2}}\cdot\left(\frac{\Delta f_{\mathrm{r}}}{f_{\mathrm{r2}}}\right)=\Delta f_{\mathrm{r}}$. To obtain reproducible waveforms across every scan, timing jitter must be suppressed; in DCS, carrier-envelope-phase jitter must also be managed. These jitters stem from instabilities in $\Delta f_{\mathrm{r}}$ and carrier-envelope offset frequency $f_{\mathrm{CEO}}$. Therefore, significant efforts have been devoted to active electronic stabilization for $f_{\mathrm{r}}, \Delta f_{\mathrm{r}}$, and $f_{\mathrm{CEO}}$ [7,13–19], as well as to stabilizing $\Delta f_{\mathrm{r}}$ using cavity-sharing or environment-shared dual-comb lasers [20–26].

To mitigate residual jitter, trigger processing using such as interference or nonlinear signals has been widely used [4,13,14,16,18,27–31]. Furthermore, real-time or post-processing corrections implemented via unequally spaced sample timings [32–34], field programmable gate arrays (FPGAs) [35,36], or software-based approaches [25,26,29,37–45] have enabled coherent averaging in combination with laser stabilization. Some of these methods even allow the use of free-running lasers [25,26,32,33,43–45]. Recently, we developed a software-based jitter-correction method for ASOPS using free-running lasers without an additional CW laser, namely, jitter-correcting (JC-) ASOPS [44,45]. However, experimental evaluations of correction efficiency are strongly affected by the specific phase noise profile of the lasers and the frequency setup for each measurement [46], making cross-system comparisons difficult. Furthermore, a quantitative theoretical framework that directly links laser phase noise spectra to jitter in signal and spectral power degradation across both ASOPS and DCS remains lacking.

To address this gap, this work establishes a comprehensive analytical and simulation framework that directly bridges laser phase noise spectra to the performance of jitter-suppression methods in ASOPS. The key contributions of this paper are threefold. First, we mathematically formulate the offset-frequency-dependent jitter suppression dynamics, explicitly revealing the significance of the software-based calibration performance over traditional trigger processing. Second, our propagation model enables the quantitative prediction of residual jitter prior to hardware construction, eliminating the need for trial-and-error system assembly. Third, by formulating the relation between spectral power degradation and RMS jitter, we introduce a powerful compensation scheme that successfully retrieves a true "jitter-free" spectrum from experimental data obtained using free-running lasers.

The remainder of this paper is organized as follows. Section 2 formulates the phase-noise suppression models for both triggering and jitter-correction methods, illustrating their low-offset-frequency scaling behaviors. Section 3 simulates noise propagation from readily measurable laser repetition frequencies $f_{\mathrm{r1}}$ and $f_{\mathrm{r2}}$ to ASOPS signals after jitter correction, and evaluates the performance of JC-ASOPS. Section 4 derives the spectral power degradation formula and demonstrates the experimental retrieval of a jitter-free spectrum in terahertz time-domain spectroscopy. Overall, this framework serves as a universal design and evaluation tool for optimizing laser selection, electronic processing, and improving spectral consistency for next-generation dual-comb and ASOPS applications.

## 2. Analytical model of triggering and jitter correction methods

To improve the signal-to-noise ratio (SNR) of the ASOPS or DCS measurements, time-domain data recurring at $\frac{1}{\Delta f_{\mathrm{r}}}$, referred to here as ASOPS and DCS signals, are typically averaged. Because timing jitter induces time-axis distortions and phase misalignments across consecutive scans, these jittered signals must undergo specific processing prior to averaging to preserve signal coherence. Most commonly, this is performed using trigger processing. In this section, we construct a unified mathematical model to directly compare the phase-noise suppression mechanisms of traditional trigger processing and software-based calibration, such as jitter correction (JC-ASOPS).

As shown in Fig. 1(a), both approaches utilize periodic reference events occurring at the same rate as the ASOPS signal, $\Delta f_{\mathrm{r}}$ (or its integer harmonics $n\Delta f_{\mathrm{r}}$). In practice, trigger or calibration signals used to define these events are generated via optical interference, nonlinear mixing, or electrical RF beating between the two lasers. For a generalized theoretical comparison, we treat trigger and calibration signals as equivalent timing references. In trigger processing, each trigger event is used as an independent time origin to realign the signal. Since the trigger signal and the ASOPS signal exhibit nearly identical jitter, the ASOPS jitter is reset to zero at each time origin, as illustrated in Fig. 1(b). As a result, the SNR near the time origin increases through averaging of the aligned signals, whereas the signal power at points far from the origin decreases because of residual jitter accumulated over time. JC-ASOPS is conceived to overcome these inherent limitations in trigger-based methods. In JC-ASOPS, consecutive calibration events define both ends of each signal segment prior to averaging, thereby calibrating the telescopic time-axis distortions caused by fluctuation in $\Delta f_{\mathrm{r}}$, as shown in Fig. 1(c). By suppressing the jitter continuously over the entire signal duration rather than just resetting it at discrete origins, the power reduction during coherent averaging is mitigated, leading to an extended measurement bandwidth.

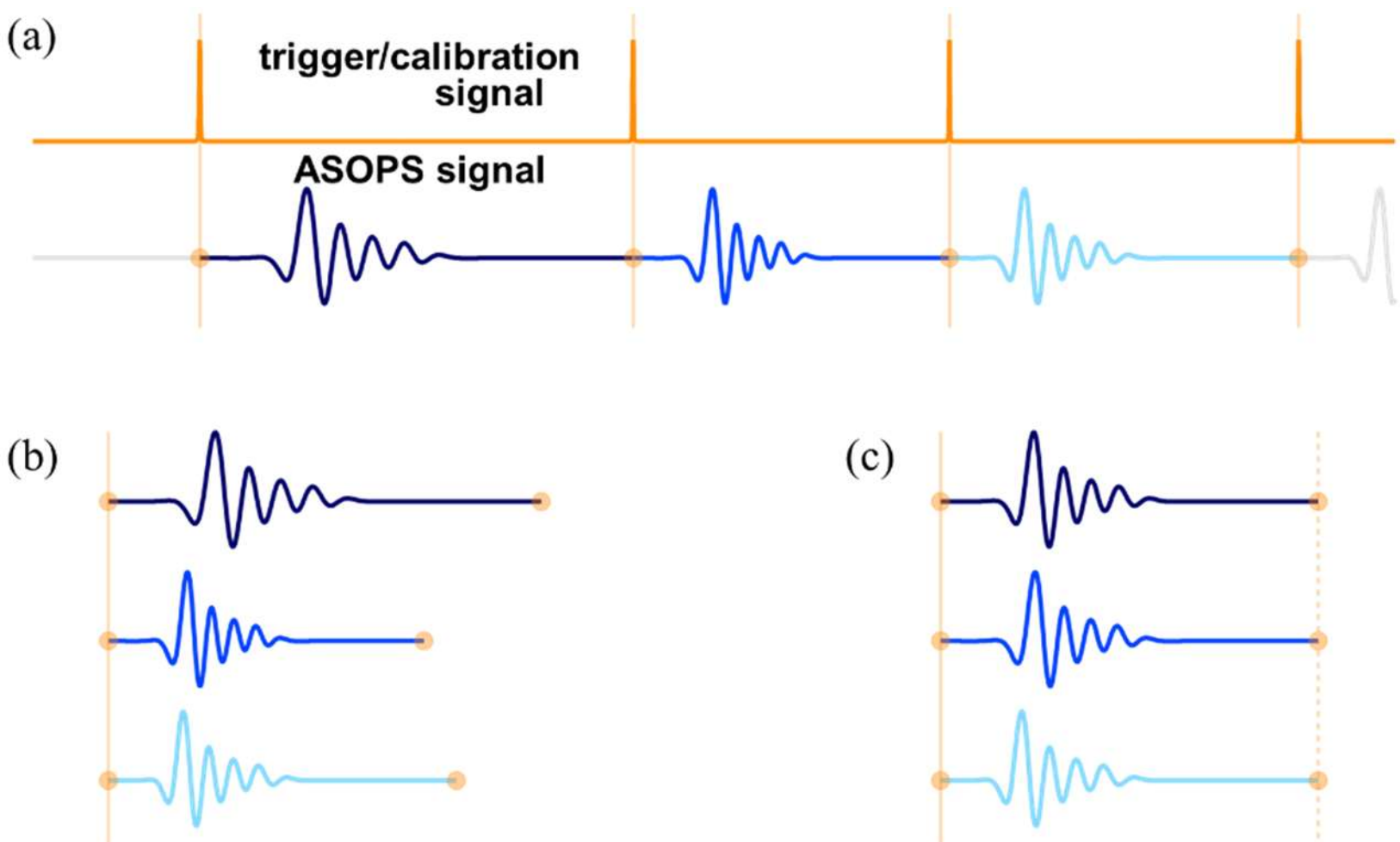

Fig. 1. Jitter suppression concept when using triggers and calibration in jitter correction methods. (a) Relation between jittered ASOPS signals and trigger/calibration signals; (b) aligned ASOPS signal using trigger; and (c) calibrated ASOPS signal in jitter correction methods.

To clarify the jitter-suppression effects of trigger processing and calibration in JC-ASOPS, we evaluate the reduction of phase noise, which provides a convenient means of describing the frequency dependence of timing jitter. When a signal with carrier frequency $f_{\mathrm{c}}$ and phase noise $\phi(t)$ is expressed as $V\cos\left(2\pi f_{\mathrm{c}}t+\phi(t)\right)$, the phase noise is typically characterized by its one-sided power spectral density $S_{\phi}(f_{\mathrm{o}})$, where $f_{\mathrm{o}}$ is the frequency of phase noise, which is generally called offset frequency. To evaluate the suppression at a specific offset frequency $f_{\mathrm{o}}$, we define the phase-noise amplitude-density component per unit bandwidth as

$$\tilde{\phi}_{f_{\mathrm{o}}}(t)=\sqrt{2S_{\phi}(f_{\mathrm{o}})}\sin\left(2\pi f_{\mathrm{o}}t+\theta(f_{\mathrm{o}})\right), \tag{1}$$

where $\tilde{\phi}_{f_{\mathrm{o}}}(t)$ has units of rad/$\sqrt{\mathrm{Hz}}$, and $\theta(f_{\mathrm{o}})$ is the phase of the component. For simplicity, we set $\theta(f_{\mathrm{o}})=0$ in the following analysis; the resulting initial phase noise is shown by the black line in Fig. 2. Figure 2(a) illustrates the phase noise component to be subtracted in each method, shown relative to the initial component (black line), where the repetition rate of the trigger or calibration signal is 10-times higher than $f_{\mathrm{o}}$. Figure 2 (b) shows the residual phase-noise component obtained by subtracting the correction term in (a) from the initial component. In the trigger processing case (yellow lines), the phase noise at each trigger event is subtracted as a step function for the corresponding signal segment. As a result, the residual phase noise is reset to zero at each trigger event and tends to increase with time away from the trigger, as shown by the yellow lines in Fig. 2. In contrast, the calibration in JC-ASOPS (blue lines) suppresses the phase noise not only at calibration events but also between them, by assuming that the phase noise varies linearly with time. Consequently, while both triggering methods and JC-ASOPS reduce the mean power of phase noise, the residual jitter for JC-ASOPS is substantially smaller than that in triggering methods, as shown in Fig. 2(b). Moreover, the time-domain behavior described above indicates that the mean power of residual jitter gets smaller when a trigger or calibration signal with a higher frequency is used.

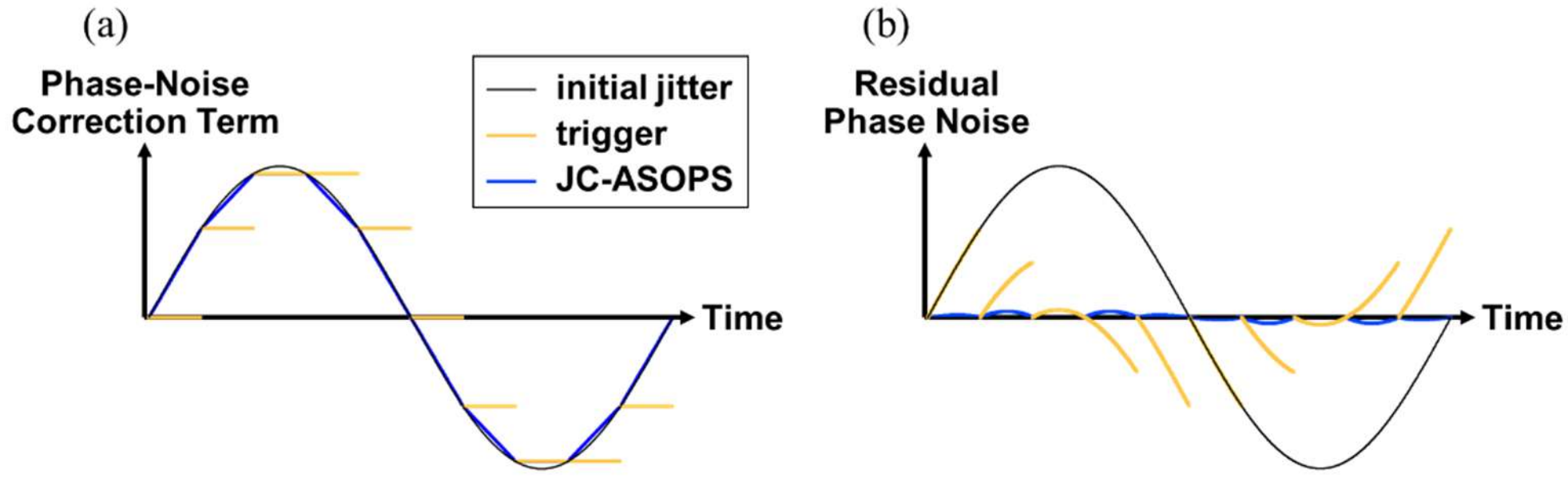

Fig. 2. Jitter suppression effects calculated for phase noise on single offset frequencies. (a) Time-domain representation of the phase-noise correction term in each method, shown relative to the initial phase noise (black line). The yellow and blue lines indicate trigger processing and JC-ASOPS calibration, respectively. (b) Residual jitter after subtracting the correction terms in (a). (c) and (d) Phase noise PSD ratios after applying each method, normalized to the initial power and shown on linear and logarithmic scales, respectively, as function of $\frac{f_\mathrm{o}}{f_\mathrm{t}}$.

Next, we investigate the offset frequency dependence of jitter-suppression effects using analytical formulae for the same model. In this analysis, the offset frequencies are assumed to be sufficiently low compared with the effective bandwidths of the trigger and calibration signal processing. In addition, the changes in phase noise power discussed above are approximated to affect only the initial offset frequencies, while power transfer between different offset frequencies induced by trigger or jitter correction is neglected. This approximation is also reasonable for low offset frequencies. For comparison with the residual phase-noise after trigger and jitter correction, we first express initial phase-noise power spectral density (PSD) at $f_\mathrm{o}$, in a time-averaged form,

$$S_\phi(f_\mathrm{o}) = \frac{1}{2\pi}\int_0^{2\pi} \tilde{\phi}_{f_\mathrm{o}}^2(a)\,da = \frac{2S_\phi(f_\mathrm{o})}{2\pi}\int_0^{2\pi} \sin^2 a\,da, \tag{2}$$

where $a \equiv 2\pi f_\mathrm{o} t$. When the trigger processing is applied, the instantaneous phase noise for the immediately preceding trigger is always subtracted from the current phase noise as discussed. Therefore, the subtraction term can be expressed as $\sqrt{2S_\phi}\sin(a - 2\pi f_\mathrm{o}\tau)$, where $\tau$ is the time elapsed from the immediately preceding trigger. Accordingly, the residual phase-noise PSD is obtained as a sinc function:

$$\begin{aligned} S_{\phi,\mathrm{trig}} &= \frac{2S_\phi(f_\mathrm{o})}{4\pi^2 \frac{f_\mathrm{o}}{f_\mathrm{t}}}\int_0^{2\pi}\int_0^{2\pi\frac{f_\mathrm{o}}{f_\mathrm{t}}} (\sin a - \sin(a-b))^2\,db\,da \\ &= 2S_\phi(f_\mathrm{o})\left(1 - \frac{\sin\left(2\pi\frac{f_\mathrm{o}}{f_\mathrm{t}}\right)}{2\pi\frac{f_\mathrm{o}}{f_\mathrm{t}}}\right) \\ &\cong \frac{4\pi^2}{3} S_\phi(f_\mathrm{o})\left(\frac{f_\mathrm{o}}{f_\mathrm{t}}\right)^2, \qquad \frac{f_\mathrm{o}}{f_\mathrm{t}} \ll 1, \end{aligned} \tag{3}$$

where $b \equiv 2\pi f_\mathrm{o}\tau$, and $f_\mathrm{t}$ denotes the trigger event frequency. For JC-ASOPS, the residual phase-noise PSD is obtained by applying linear subtraction slope between consecutive calibration events, yielding:

$$
\begin{aligned}
S_{\phi,\mathrm{JC}} &= \frac{2S_\phi(f_\mathrm{o})}{4\pi^2\frac{f_\mathrm{o}}{f_\mathrm{t}}}\int_0^{2\pi}\int_0^{2\pi\frac{f_\mathrm{o}}{f_\mathrm{t}}}\left(\sin a \right.\\
&\left. -\frac{\left(2\pi\frac{f_\mathrm{o}}{f_\mathrm{t}}-b\right)\cdot\sin(a-b)+b\cdot\sin\left(a-b+2\pi\frac{f_\mathrm{o}}{f_\mathrm{t}}\right)}{2\pi\frac{f_\mathrm{o}}{f_\mathrm{t}}}\right)^2 db\,da\\
&= 2S_\phi(f_\mathrm{o})\left(\frac{5\left(\pi\frac{f_\mathrm{o}}{f_\mathrm{t}}\right)^2-3+\left[3+\left(\pi\frac{f_\mathrm{o}}{f_\mathrm{t}}\right)^2\right]\cdot\cos\left(2\pi\frac{f_\mathrm{o}}{f_\mathrm{t}}\right)}{6\left(\pi\frac{f_\mathrm{o}}{f_\mathrm{t}}\right)^2}\right)\\
&\cong \frac{2\pi^4}{15}S_\phi(f_\mathrm{o})\left(\frac{f_\mathrm{o}}{f_\mathrm{t}}\right)^4 \quad ,\frac{f_\mathrm{o}}{f_\mathrm{t}}\ll 1,
\end{aligned}
\tag{4}
$$

where $f_\mathrm{t}$ represents the calibration event frequency. Both trigger processing (yellow) and JC-ASOPS (blue) suppress jitter more effectively at lower $f_\mathrm{o}$ and higher $f_\mathrm{t}$, consistent with the well-known experimental observation that a jitter gets smaller under higher $\Delta f_\mathrm{r}$ when $f_\mathrm{t}=\Delta f_\mathrm{r}$. More importantly, the mathematical formulation explicitly reveals distinct low-frequency scaling behavior of the reduction: $\sim\left(\frac{f_\mathrm{o}}{f_\mathrm{t}}\right)^2$ for the triggering methods and $\sim\left(\frac{f_\mathrm{o}}{f_\mathrm{t}}\right)^4$ for the jitter correction when $\frac{f_\mathrm{o}}{f_\mathrm{t}}$ is sufficiently small. This difference indicates that jitter correction provides substantially stronger suppression than trigger processing in the low-offset-frequency region.

## 3. Simulation of phase noise propagation

In this section, the phase noise in the ASOPS signal, as well as its suppression, is simulated directly from the measured phase noise in $f_\mathrm{r}$. By processing the simulated signal with our experimental JC-ASOPS program [44,45], additional effects neglected in the simplified analytical model of Section 2 are naturally included, such as noise transfer across different offset frequencies, filtering effect to reduce measurement noise in the calibration signal, and imperfections in zero-crossing detection.

### *3.1 Laser phase noise on ASOPS signal*

Since the repetition frequency of the down-converted ASOPS signal is $\Delta f_\mathrm{r}=f_{\mathrm{r}2}-f_{\mathrm{r}1}$, its phase noise directly reflects the relative phase fluctuations between $f_{\mathrm{r}2}$ and $f_{\mathrm{r}1}$. Accordingly, assuming that the two lasers are independent, the total phase-noise power is obtained by summing the individual powers. Fig. 3(a) shows the measured phase-noise PSD in $f_\mathrm{r}$ of the lasers used for the ASOPS measurement. The summed phase-noise PSD shown in Fig. 3(b) is used as input in the simulation. The phase-noise PSD outside the measurement range is assumed to be flat, matching the nearest measured boundary values, and the phase at each offset frequency was assumed to be random. To simulate the fundamental wave of the down-converted time-domain signals, the IFFT of the complex phase noise spectrum, extended to negative offset frequencies using complex-conjugate symmetry, is used as $\phi(t)$ in the following formula:

$$V_1(t)=V_{0,1}\exp\left(i2\pi\Delta f_\mathrm{r}t+i\phi(t)\right). \tag{5}$$

The phase-noise PSD recalculated from the simulated waveform is shown by the black line in Fig. 3(b). The $n$th-order harmonic components of both the ASOPS and calibration signals, with frequency $n\Delta f_\mathrm{r}$, are similarly expressed as

$$V_n(t)=V_{0,n}\exp\left(i2\pi n\Delta f_\mathrm{r}t+in\phi(t)\right). \tag{6}$$

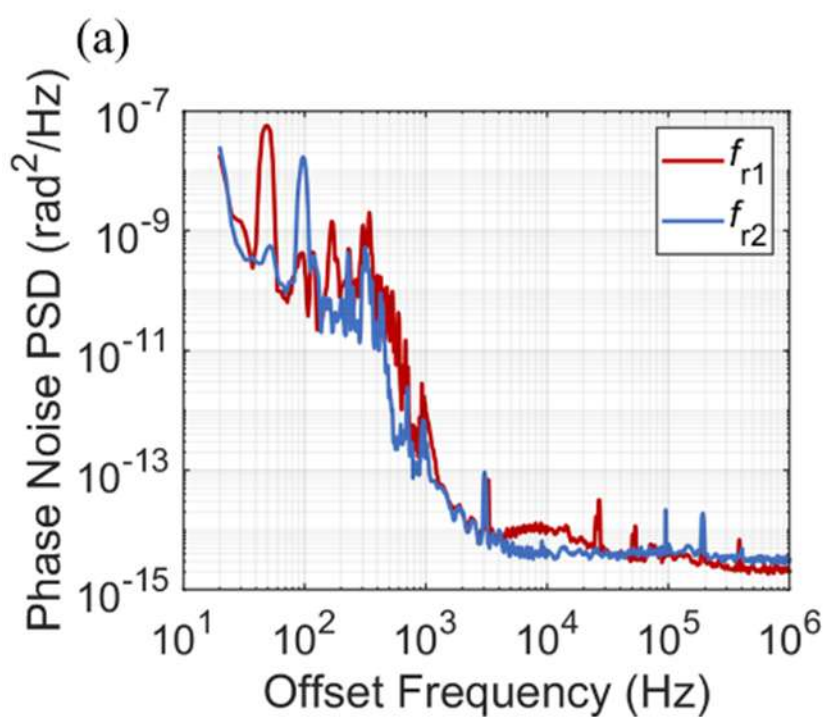

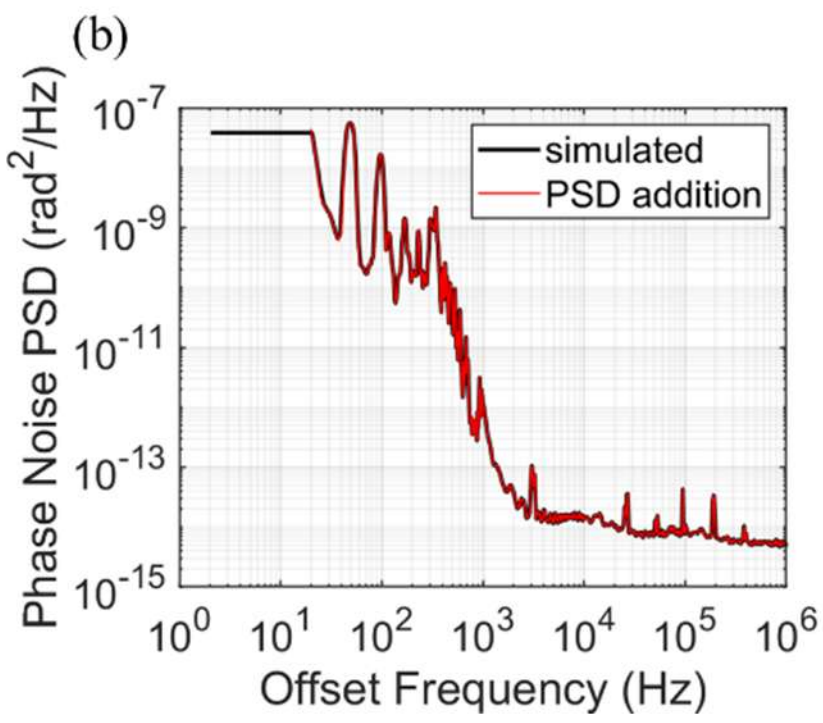

Fig. 3. Simulation of the phase noise in $\Delta f_{\mathrm{r}}$. (a) Measured phase noise power spectrum density (PSD) in laser $f_{\mathrm{r}}$; (b) phase noise PSD in $\Delta f_{\mathrm{r}}$ calculated by power addition (red), and the phase noise in the wave simulated by using the PSD addition (black).

*3.2 Phase noise suppression by jitter correction*

The jitter suppression effect achieved by arbitrary ASOPS calibration software can be simulated by replacing the raw experimental data with the synthetic waveforms calculated from Eqs. (5) and (6). We simulated the phase-noise suppression of our JC-ASOPS program for the fundamental components of the ASOPS signals generated from Eq. (5). In experimental studies, selecting appropriate values of calibration-signal frequency and $\Delta f_{\mathrm{r}}$ is often important and challenging. Furthermore, $\Delta f_{\mathrm{r}}$ varies during the measurement when free-running lasers are used. Therefore, the phase-noise suppression was evaluated for different values of $\Delta f_{\mathrm{r}}$ and calibration-signal frequencies, $n\Delta f_{\mathrm{r}}$. In the jitter-correction program, the zero-crossing times of the calibration signals are determined after filtering and used to calibrate the measured time axes [45].

The solid lines in Fig. 4(a)–(f) show the phase noise in the ASOPS signals before (black) and after (blue) the jitter correction. The dashed black lines indicate the calibration signal frequency used in each simulation for reference. In the simulations using low-frequency calibration signals, as shown in Fig. 4(a) and (b), the phase noise was mainly reduced at offset frequencies below the calibration signal frequencies, with greater suppression at lower offset frequencies, consistent with the prediction in section 2. When the analytical calculation predicts very low residual noise below the calibration-signal frequency, however, the phase noise after jitter correction converged to a floor of $\sim 10^{-12}$. This tendency is particularly noticeable at small offset frequencies, where the phase-noise suppression is greatest, or when the initial phase noise below the calibration signal frequency is already low. These results are attributed to imperfect calibration resulting from digital filtering and smoothing applied to suppress amplitude noise during zero-crossing detection.

The residual jitter estimated by integrating the simulated phase noise under different conditions is shown in Fig. 4(g). The integration bandwidth was set to 1 MHz, considering the noise floor of the phase noise measurement. The yellow plots show the residual jitter for $\Delta f_{\mathrm{r}} = 100\ \mathrm{Hz}$ with varying harmonic order $n$ of calibration signals, whereas the gray plots show the residual jitter for $n = 20$ with varying $\Delta f_{\mathrm{r}}$. Except for slight discrepancies caused by the simulation resolution and random fluctuations, the two curves exhibit almost identical dependence on the effective calibration signal frequency, $n|\Delta f_{\mathrm{r}}|$. Crucially, this simulation reveals that increasing the frequency of the calibration signal yields rapid jitter reduction initially, but does not endlessly improve performance; an optimal frequency exists (~2 kHz in this system), beyond which performance stabilizes. This trend is consistent with our experimental observations.

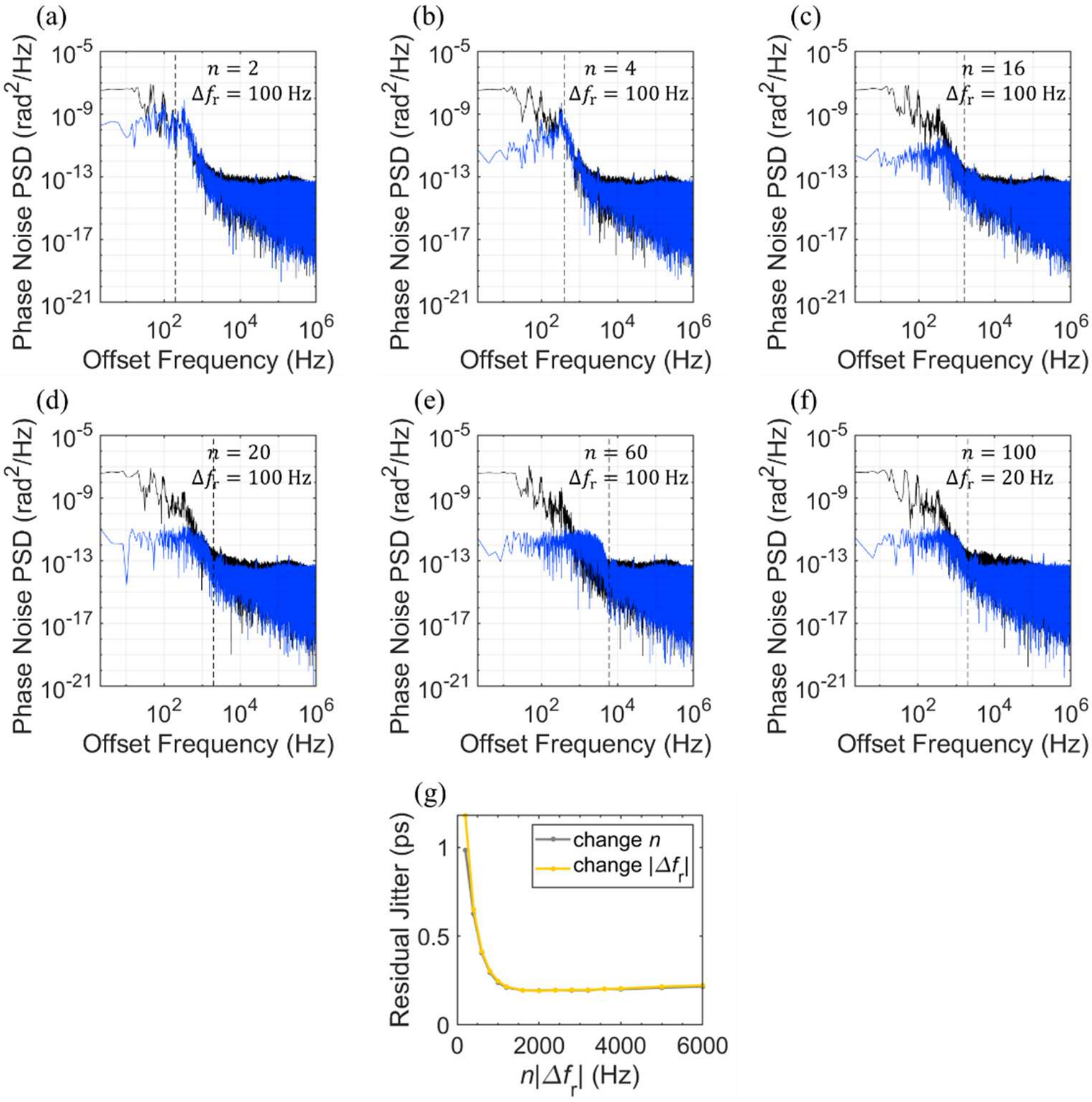

Fig. 4. Simulated effects of jitter correction. (a)–(f) Simulated phase noise before (black) and after (blue) the jitter correction, with the calibration signal specified by the inset expression and its frequency indicated by the vertical dotted lines; (g) residual jitter for different calibration signal parameters.

## 4. Calculation and compensation of spectral power degradation

In most spectroscopic applications, including ASOPS, the quantitative stability and fidelity of the resulting spectra are essential for accurate comparison. However, timing jitter in time-domain waveforms induces phase misalignments, which inevitably reduces the spectral power upon coherent averaging. Particularly when using free-running lasers, the degree of degradation can fluctuate significantly depending on the jitter level with varying $\Delta f_{\mathrm{r}}$. Fortunately, both the initial and residual jitter level can be directly estimated from measured experimental data in JC-ASOPS [45]. In this section, we derive a general analytical formula to calculate the power reduction from the jitter level, and demonstrate the experimental retrieval of a true jitter-free spectrum.

When the averaging interval $\Delta t$ is an integer multiple of $\frac{1}{\Delta f_{\mathrm{r}}}$, the scan integration acts as a coherent averaging for ASOPS signals, because each jitter-free frequency component in ASOPS signal satisfies

$$\exp(i2\pi N\Delta f_{\mathrm{r}} t) = \exp\big(i2\pi N\Delta f_{\mathrm{r}}(t+\Delta t)\big), \tag{7}$$

where $N$ denotes the index of the frequency component in the ASOPS signal. Accordingly, the average of the ASOPS signal with phase noise over $A$ consecutive scans can be written as:

$$\frac{1}{A}\sum_{a=0}^{A-1}\sum_{N} V_N(t+a\Delta t)=\frac{1}{A}\sum_{a=0}^{A-1}\sum_{N} V_{0,N}\exp(i2\pi N\Delta f_{\mathrm{r}}t)\cdot\exp\big(in\phi(t+a\Delta t)\big). \tag{8}$$

The spectrum power of each frequency component is then given by

$$\left|\frac{1}{A}\sum_{a=0}^{A-1} V_{0,N}\exp(i2\pi N\Delta f_{\mathrm{r}}t)\cdot\exp\big(in\phi(t+a\Delta t)\big)\right|^2$$
$$=\frac{1}{A^2}V_{0,N}^2\sum_{a=0}^{A-1}\sum_{a'=0}^{A-1}\exp(iN[\phi(t+a\Delta t)-\phi(t+a'\Delta t)]). \tag{9}$$

Therefore, the expected mean power after averaging becomes [47]

$$\mathbb{E}\left(\frac{V_N^2}{2}\right)=\frac{1}{A^2}V_{0,N}^2\sum_{a=0}^{A-1}\sum_{a'=0}^{A-1}\mathbb{E}[\exp(iN[\phi(t+a\Delta t)-\phi(t+a'\Delta t)])]$$
$$=V_{0,N}^2\,\mathbb{E}\big[\exp\big(iN\phi(t)\big)\big]\cdot\mathbb{E}^*\big[\exp\big(iN\phi(t)\big)\big]=\lim_{T\to\infty}\frac{V_{N,0}^2}{2T^2}\left|\int_{-\frac{T}{2}}^{\frac{T}{2}}\exp\big(iN\phi(t)\big)\,dt\right|^2. \tag{10}$$

When the phase noise is represented as a Fourier expansion of offset-frequency components with random phases,

$$\phi(t)=\sum_k\sqrt{2S_{\phi,k}}\cos\big(2\pi f_{\mathrm{o},k}t+\theta_k\big), \tag{11}$$

Eq. (10) can be written as,

$$\mathbb{E}\left(\frac{V_N^2}{2}\right)=\lim_{T\to\infty}\frac{V_{N,0}^2}{2T^2}\left|\int_{-\frac{T}{2}}^{\frac{T}{2}}\prod_k\exp\left(iN\sqrt{2S_{\phi,k}}\cos\big(2\pi f_{\mathrm{o},k}t+\theta_k\big)\right)dt\right|^2$$
$$=\lim_{T\to\infty}\frac{V_{N,0}^2}{2T^2}\left|\int_{-\frac{T}{2}}^{\frac{T}{2}}\prod_k J_0\left(N\sqrt{2S_{\phi,k}}\right)dt\right|^2=\frac{V_{N,0}^2}{2}\left[\prod_k J_0\left(N\sqrt{2S_{\phi,k}}\right)\right]^2, \tag{12}$$

where $J_0(z)$ is the zero-th order Bessel function of the first kind, satisfying $\frac{1}{\pi}\int_{-\pi}^{\pi}\exp(iz\cos\vartheta)\,d\vartheta=J_0(z)$. For sufficiently small amplitudes, $\sqrt{2S_{\phi,k}}\ll 1$, Eq. (12) can be approximated as

$$\mathbb{E}\left(\frac{V_N^2}{2}\right)=\frac{V_{N,0}^2}{2}\exp\left\{\sum_k\ln\left[J_0\left(N\sqrt{2S_{\phi,k}}\right)\right]\right\}^2$$
$$\approx\frac{V_{N,0}^2}{2}\exp\left\{\sum_k-\frac{1}{4}\left(N\sqrt{2S_{\phi,k}}\right)^2\right\}^2=\frac{V_{N,0}^2}{2}\exp\left(-N^2\sum_k S_{\phi,k}\right)=\frac{V_{N,0}^2}{2}\exp\big(-N^2\phi'^2\big), \tag{13}$$

where $\phi'$ represents the root-mean-square (RMS) jitter. Crucially, Eq. (13) demonstrates that, when the phase noise is random and sufficiently small, the power reduction in ASOPS spectra is determined solely by the RMS jitter $\phi'$, regardless of the shape of the phase-noise spectrum. Furthermore, the power reduction factor relative to the initial power, $\frac{V_{N,0}^2}{2}$, follows a Gaussian function centered at 0 Hz as a function of the ASOPS signal frequency, $N\Delta f_{\mathrm{r}}$ (or the corresponding terahertz frequency, $Nf_{\mathrm{r}}$).

To experimentally validate this power compensation scheme, we applied Eq. (13) to terahertz time-domain spectroscopy (THz-TDS) performed via JC-ASOPS with two free-running lasers, as illustrated in Fig. 5(a). For comparison, a mechanical scan (standard THz-TDS) using a delay line was also conducted using a single laser (the same source used for terahertz generation in ASOPS). The optical components and paths used in the two measurements are kept essentially identical, except for the gating optics indicated by dotted lines in Fig. 5(a). To retrieve the true jitter-free spectrum from the JC-ASOPS measurement, the spectral power was divided by the Gaussian factor calculated using the experimentally measured residual RMS jitter, 0.14 ps. Figure 5(b) presents the measured (blue) and retrieved (light blue) terahertz spectra obtained by JC-ASOPS. The pink curve represents the spectrum obtained via the mechanical scan, scaled by a constant factor to match the vertical offset differences in detector sensitivities. While dividing by the Gaussian profile elevates the high-frequency noise floor into a rising baseline, the retrieved spectrum demonstrates remarkably excellent agreement with the mechanical scan spectrum across the primary optical bandwidth. Because mechanical delay scans are fundamentally immune to multi-scan timing jitter, this striking agreement confirms that our mathematical formulation successfully compensates for jitter-induced power degradation, providing a powerful post-processing tool to reconstruct pristine, highly consistent spectra from free-running ASOPS/dual-comb measurements.

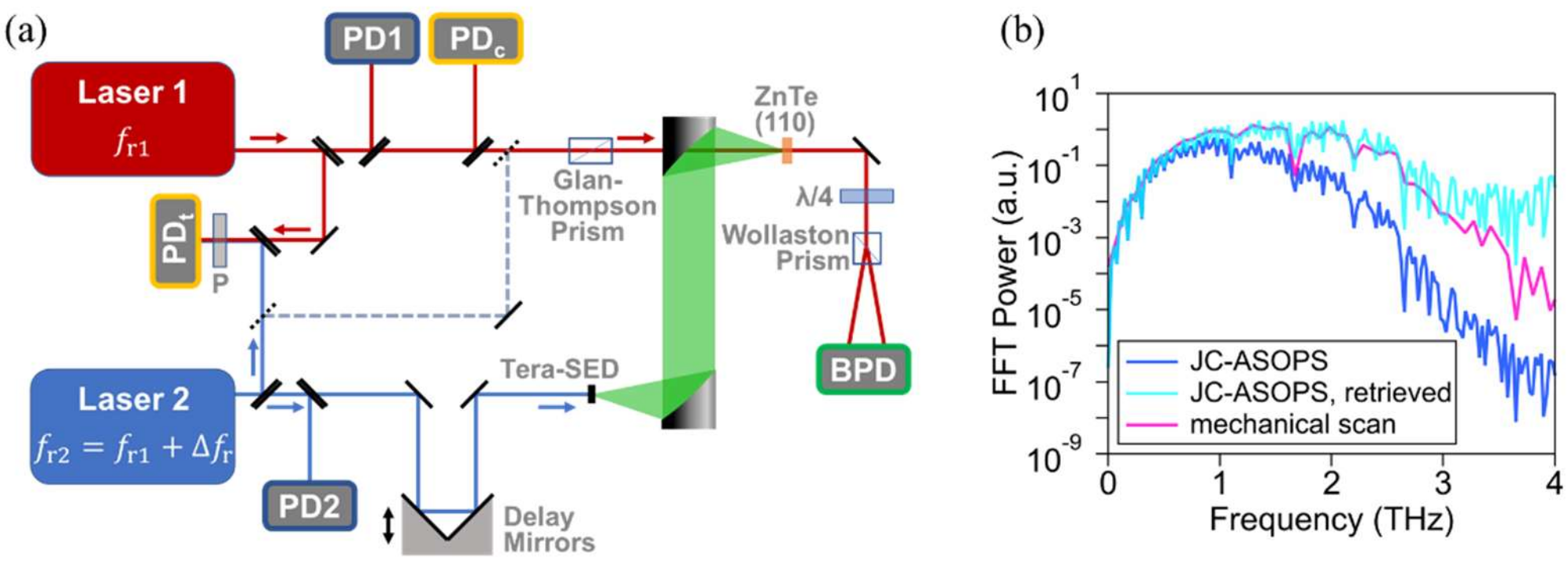

Fig. 5. Power spectrum retrieval. (a) Experimental setup for comparing THz-TDS power spectra obtained by JC-ASOPS and mechanical scan (with mirrors shown by dotted lines); (b) Power spectrum measured by JC-ASOPS (blue), retrieved spectrum from the JC-ASOPS measurement using measured residual jitter (light blue), and power spectrum measured by mechanical scan (pink).

## 5. Conclusion

In conclusion, we have developed a unified analytical framework to model and compensate for phase noise and timing jitter propagation in ASOPS and DCS. Our analytical model and simulations revealed that residual phase noise scale as $\left(\frac{f_o}{f_t}\right)^2$ through triggering methods and as $\left(\frac{f_o}{f_t}\right)^4$ through the jitter correction method, in the low-offset-frequency region. The phase noise after applying the experimental jitter-correction program can be simulated from measured phase noise of the laser repetition frequency, $f_r$, enabling jitter predictions prior to constructing an ASOPS system. The simulated jitter varied with the calibration-signal frequency, indicating the existence of an optimal frequency. Notably, the proposed simulation framework can readily be applied to ASOPS jitter-correction methods in quite general. Furthermore, we established that spectral power degradation upon coherent averaging follows a universal Gaussian decay governed solely by the residual RMS jitter, independent of the phase noise spectral shape. This theoretical framework was experimentally validated using a THz-TDS setup via JC-ASOPS. By compensating for the spectral power degradation, the retrieved spectrum agreed exceptionally well with a reference mechanical scan spectrum. These findings offer essential design guidelines and practical tools for high-precision ASOPS and DCS with free-running lasers.

**Funding.** Japan Society for the Promotion of Science (JSPS) KAKENHI (JP22K18269).
**Disclosures.** The authors declare no conflicts of interest.
**Data availability.** The data for this study are available from the corresponding authors upon request.